**Graph-Based Recommendation System Enhanced by Community Detection**

**Zeinab Shokrzadeh, Department of Computer Engineering, Urmia Branch, Islamic Azad University, Urmia, Iran. (Email: Shokrzadeh82@yahoo.com)**

**Mohammad-Reza Feizi-Derakhshi, Department of Computer Engineering, University of Tabriz, Tabriz, Iran. (Email: mfeizi@tabrizu.ac.ir)**

**Mohammad-Ali Balafar, Department of Computer Engineering, University of Tabriz, Tabriz, Iran. (Email: balafarila@tabrizu.ac.ir)**

**Jamshid Bagherzadeh Mohasefi, Department of Computer Engineering, Urmia University, Urmia, Iran. (Email: J.bagherzadeh@urmia.ac.ir)**

**Abstract**

Many researchers have used tag information to improve the performance of recommendation techniques in recommender systems. Examining the tags of users will help to get their interests and leads to more accuracy in the recommendations. Since user-defined tags are chosen freely and without any restrictions, problems arise in determining their exact meaning and the similarity of tags. However, using thesaurus and ontologies to find the meaning of tags is not very efficient due to their free definition by users and the use of different languages in many data sets. Therefore, this article uses mathematical and statistical methods to determine lexical similarity and co-occurrence tags solution to assign semantic similarity. On the other hand, due to the change of users' interests over time this article has considered the time of tag assignments in co-occurrence tags for determining similarity of tags. Then the graph is created based on similarity of tags. For modeling the interests of the users, the communities of tags are determined by using community detection methods. So, recommendations based on the communities of tags and similarity between resources are done. The performance of the proposed method has been evaluated using two criteria of precision and recall through evaluations on two public datasets. The evaluation results show that the precision and recall of the proposed method have significantly improved, compared to the other methods. According to the experimental results, the criteria of recall and precision have been improved, on average by 5% and 7% respectively.

## 1. Introduction

People face a rapid and huge growth of data in social systems. Although there is a lot of useful information in various fields, finding accurate and desirable data is difficult and time-consuming. To conquer this problem, recommender systems have been provided. Those systems are software techniques and tools that assist users in various decision-making processes. In fact, while users need to find the right information, they need a system that supports them. One of the offered solutions in this field is the development of recommender systems to provide personalized services according to users' interests. Recommender systems are used in various fields and applications. One of the most popular well-known systems implemented is Amazon website, which takes advantage of customers' purchase behavior, attractions, and offers according to the users' interests.

The overall structure of a recommender system follows a set of phases including collection, learning, and recommendation[1][2]. In the first phase, appropriate resources that comprise the relevant information of users are selected. Then, a leaner (supervised or unsupervised learning) analyzes the users' preferences and extracts their behavioral patterns. The final phase recommends the entities that are the most similar to the users' interests. It is important to recognize that, within a common core structure of a recommender system, there are variations from application to application. Some of the most sophisticated and heavily used recommender systems in the industry are Last.fm, YouTube, and Amazon [3].

Generally, recommender systems can generate a list of recommendations by these approaches: content-based filtering, collaborative filtering, hybrid recommender systems and, so on [4]. Based on the existing research, the conventional $CF^1$ approaches, which only use user-item rating information to make recommendations, are in two major categories: the memory-based CF and model-based CF both of which can be used to make recommendations in tagging systems [5]. Memory-based methods make suggestions based on the nearest neighbors and model-based recommendation based on the model created by users. In recommender systems, another type of system was introduced as tagged or tagging systems. Mathes discussed the tags on the web in articles in late 2004 [6].

Recently, social tagging systems have become an important instrument of Web 2.0 that allows users to collaboratively annotate and search the content [7] [8]. To facilitate this process, the present research has attempted to improve the performance and quality of resource recommendations. Despite the creation of new opportunities, social tagging recommender systems, revive old problems such as information overload. Recommender systems good applications in making available the information that is related to the users' interests. However, we face new challenges in tagging recommender systems.

In these systems, users are interested in finding tags, contents and even other users. Furthermore, while traditional recommender systems typically work with 2D data arrays, the data in these systems act as a third-order tensor or a multilayer graph with user nodes, resources, and tags which have been introduced as new aspects of recommendations such as users, resources and introduced the tags. Therefore, new approaches and algorithms were needed to address the threefold nature of the data in these systems. Various social tagging systems such as Del.icio.us, Last.Fm, CiteULike, Flickr, and others allow users to assign custom tags for resources based on their background knowledge to manage, organize, share, discover and retrieve resources [9]. These systems aggregate the information of heterogeneous elements to have enriched information. The role of tags in the systems is essential [10].

Collaborative tagging systems, also known as folksonomy, have grown dramatically on the Web. Tags in these systems significantly organize the content of websites and other resources and effectively display user behavior. This is considered an advantage for these systems. Tags are also used as a bridge between users and resources to describe users' interest in resources [11]. Researchers use a variety of strategies to gain the users' interests and make recommendations with greater accuracy. In fact, one of the most important concerns in the field of recommender systems is to provide more accurate recommendations according to the users' interests.

This article focuses on one of the major challenges of recommender systems, which is to improve the performance of recommender algorithms. To improve the performance, it has relied on modeling user' interests and tagging clustering based on user tagging behaviors. Clustering and providing more accurate analysis we have used the community detection method. This is considered an important achievement. In fact, by examining the tagging behaviors of users more closely, using the suggested similarity criteria, forming a graph, and using the community detection method, we have paved the way to obtain users' interests and finally, we have increased the strength of the recommendations with the nearest neighborhood method.

Tagging activities in folksonomies are not guided by any formal regulations (no dictionaries, no thesaurus) meaning that users can tag resources with any tags they like [12]. This leads to a wide variety of tags like inflections, spelling errors, abbreviations creative use of compounding, and etc. We can interpret the tags in the folksonomy as concepts [13]. As a result, in tagging-based recommender systems, the main problems arise from discovering the meanings of tags. Due to the ambiguity in the meanings of the tags and lack of correct discovery of their meanings, the performance of these systems is affected. The co-occurrence tag method has been used for semantic communication in most previous studies. The number of studies, ontologies or external knowledge have been used to strengthen this method. In semantic theory, in order to find the relationship of tags based on external knowledge, an attempt is made to adapt them with meaning. In [14] they have tried WordNet concepts for deriving relationships. However, WordNet is a static resource, and only less than half (48.7%) of tags can match the direct study in [10] and described a domain ontology development approach that extracts domain terms from folksonomies and enrich them with data and vocabularies from the Linked Open Data cloud. As a result, this article obtains lightweight domain ontologies that combine the emergent knowledge of social tagging systems with formal knowledge from ontologies [15]. In general, it is difficult to choose the right concept that matches the tag due to the lack of tagging context. This is because the process of tagging users is very different from the lexicologists or domain specialists. This problem of separating the concept of tag is discussed in [15] [16]. Even if a tag can be lexically consistent with a concept in external sources, the conformity of their intended meanings is unclear [17].

---

[1] Collaborative Filtering

In this work, we presented a new method of collaborative filtering resource recommendation systems called social collaborative based on community detection with semantic and lexical connections of tags. It should be noted that when users employ tags for resources, these tags clearly show their preference and interest. By examining the interactions between users and tags, it is possible to understand the semantic correlation between resources and users, and also to extract users' interests more accurately than recommending systems based on rating.

The main contribution of this paper can be explained as the following:
- ✓ To the best of our knowledge, this is the first attempt leveraging semantic and lexical similarity of tags at the same time by considering the time of tag assignments parameter to construct graph of tags .These similarities are used to obtain the association strength of the tags.
- ✓ We apply community detection methods for clustering of tags. This leads to precise modeling of the interests of the users.
- ✓ This is the first work that used Ellenberg similarity criterion in recommendation phase for resource similarity. By using this criterion, in addition to the similarity of the resources, their differences are also taken into account.
- ✓ Based on two real-word data sets, we have conducted experiments to evaluate the effectiveness of the method. The results show that the proposed method outperform than the state-of-the-art recommendation methods achieved higher.
- ✓ This work is different from the previous methods for a number of reasons, because we have not used any external linguistic resources such as Word Net or semantic resources (like ontology) and this makes the method stronger and covers most of the tags. On the other hand, WordNet or other external knowledge is maintained manually by experts and thus remains unchanged in long term. In fact, the low coverage of WordNet inevitably leads to the poor performance of the WordNet based on tag sense disambiguation methods. On the other hand, as in the previous methods, we have used the co-occurrence tag methods by considering the time of tag assignments parameter and lexical similarity to strengthen their communication.
- ✓ Another strength of the proposed method is the use of community detection method to analyze tags and find appropriate clusters of them. All these are to improve the quality of system performance.

The rest of the paper is organized as follows: Section 2 summarizes the work done in this area, Section 3 deals with the proposed method with subsections for generating tags graph, community detection, and Resources Recommendation Stage, Section 4 describes performance evaluation of proposed solution and the results of the proposed method and finally, in Section 5, we have presented the conclusions and future works.

## 2. Related works

Nowadays, tag-based recommender algorithms are evolving rapidly. In general, tag-based recommender systems provide recommends to users by analyzing tags assigned to resources. In traditional recommender systems, especially CF, only two-dimensional data were used based on user resource rating and often with a rating resource user matrix, in tagged systems, that is, collaborative tagging, another dimension of information, namely, social tags, has been used as a powerful mechanism for making more accurate suggestions.

Although some studies has been perform on tagged recommendation systems, more research is still needed since there are many challenges in these systems. Many researchers have been trying to come up with solutions for better recommendations according to users' interest. Some of these investigations have been somewhat successful, and some have been able to respond under certain conditions. In this section, we review some related studies.

Tso-Sutter et al. [17] used tag information as an additional source along with user rating information matrix in a content-based recommender system. In their work, they extended the user-item matrix to the user-item-tag matrix and used the Jaccard similarity criterion to find neighbors. However, due to the issue of tag quality, their proposed content-based method based on memory was not very successful in improving performance. Niwa et al. [18] made an effort to recommended web pages based on the analysis of tag used and degree of relationship between tags with users. However, in this works the accuracy of the recommendations was between 40% and 60% that was not a good result. The only advantage of their proposed method compared to similar methods was the reduction of complexity due to the lack of page browsing and the use of tags. Sen et al. [19] used a special tag ranking function to obtain user tag preferences. In addition, they used additional information such as search history and click streaming, which is difficult to use in real systems compared to other methods.

Some researchers have examined various aspects of tagged systems [11, 20–24]. In field of recommendations, the information contained in these systems, has shown its significance in recommending resources, tags and users [21]. Goel and Kumar [11], the efficiency of tags in organizing the items to be encoded was examined. Article has studied the reasons for the effectiveness of social tagging systems [21]. Lamere [22], the authors examined the relevance of tags in music information retrieval. Golder and Huberman [24], the authors analyzed the structure and pattern of use of social tagging systems in Del.icio.us and compared the differences between collaborative and taxonomy tags.

Xu et al. [25] used an algorithm to recommend tags using the collaborative tagging information method. Their proposed algorithm considered the tags of a large number of users in the target document and tried to minimize the recommended concept overlapping tags to increase the level of coverage of small documents. Unfortunately, this method did not cover new documents. This is important for us to analyze in terms of tags, but we seek to recommend resources, that are different.

Zhang et al. [26], instead of analyzing tags, the authors used the features of the resources being tagged and combined them with the CF method to model user interest. This method identified implicit relationships that were absent in the traditional CF method. Determining the features of the resources was one of the problems that reduced the efficiency of the system.

Wu et al. [27] proposed the tag2word model based on a content-based method for determining the semantic relationships between the tags. Their method was able to reinforce the recommendations. It would not have worked properly if the tags had been used in the content of the documentation. This problem is obvious because many of the tags had been used by users are not in the content of the resources. Therefore, this method does not apply to all types of systems. According to the authors' research, the used dataset gives better recommendations when the usage of tags in the titles or text of sources is high. This solution was presented in a content-based method in recommender systems.

de Gemmis et al. [28] combined semantic analysis of tags with a content-based approach. They were assisted in analyzing the meaning of the WordNet for disambiguation tags. In this approach, they combined the traditional content-based method with semantic analysis of tags and provided recommendations according to user interest. But the proposed method could not also be successful in disambiguating of tags.

Wartena et al. [29] used the idea of distributing co-occurring tags and proposed a tag recommendation system. In fact, they combined the CF method with the proposed idea. Their method did not succeed compared to the other methods. They proved that when the number of tags given to resources by a single user is higher, the proposed method works better.

Usually, if we want to examine the tag-based recommender systems in terms of the type of recommendation, these systems are divided into three categories: tag suggestions, resource, and user. The type of offering these systems do not really matter because all three categories make recommendations based on the tags [30].

Ignatov et al. [31] created profiles for radio stations and users from the tags of songs they listened to, and used the online release of tags to dynamically update profiles. Vall [32] and Su et al. [33] implicitly created tag-based profiles for music recommendations. Xie et al. [34] added emotions to user profiles and tagged resources. Ignacio et al. [35] proposed a way to extend user profiles and tag-based resources to build cross-domain recommendations.

In general, nowadays there are two main approaches in the field of tagged systems, which include approaches based on graph and content [36]. In the field of graph-based solutions, graph analysis methods can be used, which is one of the methods of community-based graph analysis. There is also another issue in tagged systems, and that is related to the methods of discovering and mapping meaning to the tag.

In this object, three methods have been proposed: (1) methods based on clustering, (2) methods based on ontology, and (3) hybrid methods that combine Techniques 1 and 2. Ontology-based methods are not suitable for determining the relationship between terms.

To achieve ontology-based sustainable systems, ontology building should be done by people having domain knowledge and not just by knowledge experts [37]. This is costly and time-consuming and these methods are used in the hope of solving the problem of semantic ambiguity when they could not solve the problem [38]. In addition, because these methods use external knowledge such as WordNet and Wikis, they can't completely cover the tags used and lead to increased workload without complete problem solving.

In most cluster-based methods, external knowledge sources such as WordNet and Wikis are used to determine the semantic relationships of tags as in ontology-based methods, which have the same problems mentioned in ontology-based methods in this category of solutions. A study on the Last.Fm dataset found that over 50% of the tags used were not covered by WordNet or any other traditional lexical resources [24]. Therefore, by examining the existing methods, we came to the conclusion that the simple and effective approaches many researchers use in catching semantics to folksonomy are based on mathematical and statistical formulas. Mathematical and statistical formulas play an important role. The best thing about them is that they were clear and unambiguous [39]. Therefore, using statistical and mathematical methods, the semantic and lexical relationship of tags can be determined. In the proposed solution, we did not use any external semantic sources such as ontologies or thesaurus, however we used accurate and formal methods in determining the semantic relationships of tags, which strengthen the proposed solution for managing a large number of tags in folksonomy. Because mathematical and statistical methods have good accuracy for extracting semantic and lexical relations of tags, they are suitable to be use in the proposed solution. After determining the semantic and lexical relationships of tags, we used an effective method in clustering tags called community detection methods. That is also one of the solutions in graph-based tagging systems. With community detection methods, more accurate analysis of relationships between graph elements can be provided. In general, it is possible to make more personalized suggestions in recommending systems by using community-based solutions, a good way to analyze networks. Thus, the quality of recommendations increases and this is the advantage of our proposed solution.

This work is different from the previous methods for a number of reasons, because we have not used any external linguistic resources such as WordNet or semantic resources (like ontology) and this makes our method stronger and covers most of the tags. On the other hand, WordNet or other external knowledge is maintained manually by experts and thus remains unchanged in long term. In fact, the low coverage of WordNet inevitably leads to the poor performance of the WordNet based on tag sense disambiguation methods. On the other hand, as in the previous methods, we have used the co-occurrence tag methods by considering the time of tag-assignments parameter and lexical similarity to strengthen their communication. Another strength of the proposed method is the use of community detection method to analyze tags and find appropriate clusters of them. All these are to improve the quality of system performance

## 3. Proposed method

In this Section, we examined the users' tagging behaviors that could determine their interest. To achieve this aim, we used tags for resources and categorized them, determining users' interests.

A social tagging system consists of a set of users (U), a set of tags (T), and a set of resources (R). We define these sets as follows:

$U=\{u_1, u_2, \ldots, u_n\}$ $\quad\quad$ $T=\{t_1, t_2, \ldots, t_m\}$ $\quad\quad$ $R=\{r_1, r_2, \ldots, r_k\}$

where n is the number of users, m is the number of tags, and k represents the number of resources. In these systems, a folksonomy is defined as <U, R, T, Y>, where Y is a ternary relation between them, i.e., $y \subseteq U \times T \times R$ [40]. Although there are various general datasets available for evaluating recommender algorithms, we chose Del.icio.us dataset to evaluate the work. Because the proposed method does not use any external thesaurus or ontologies, it supports other languages than English, so it is suitable for evaluating.

The proposed approach consists of two main phases. The first phase includes two steps:1) creating a graph of tags and 2) identifying communities of tags. The second phase is to make recommendation based on the communities created from the tags and available resources in each community. In the following, we will explain the phases of the proposed solution.

### 3.1. Generating tags graph

As it was previously explained, the proposed solution includes two phases. The first phase includes two stages, the first which is the formation of tag graphs. Graph nodes of tags, and the weight of its edges are determined by the amounts of lexical, semantic similarity and the time of tag assignment. For example: the weight of two tags, $t_i$ and $t_j$, is shown with w ($t_i$, $t_j$). After generating the graph, in the second stage, the tag communities are identified. In other words, the basis of this work is detecting communities of user tags and building communities of resources based on them. For each community of tags, a community of relevant resources and users are created. Finally, resource suggestions are recommended for the target user based on the probability of membership of each resource to the communities and the power of the local neighborhood. In fact, with this new method, it is possible to identify the interest of users accurately and provide precise recommendations. To create a graph, in the first phase, for determining the relationship between the tags, use their semantic and lexical similarity and the time of tag assignment. In fact, the first innovation of the proposed method is to determine the relationship of tags by a combination of semantic (considering the time of tag assignment) and lexical similarity and not by using foreign linguistic or semantic sources. As regards, social tags are very beneficial, but due to the nature of free-form tagging and the lack of explicit meaning in social tagging systems, there are many obstacles that may prevent the useful application of social tags [7].

One of the obstacles is syntactic variations. This means a word in different syntactic forms may be used in different tags. For example, one user may annotate a web resource with the tag "picture," while another user may do this work with the tag "pictures." (Other examples are "web," "web20," "acm", "acmi", "acma", and so on.) Sometimes, words with the same meaning but very close syntax from different languages are used (e.g., "centre" and "center," "absurd" and "absurde"). These changes must be considered to provide satisfactory performance; otherwise, they may lead to confusion [7]. These are the reasons that motivated the use of lexical similarity. Semantic and lexical similarities are used to obtain weights assigned to graph nodes (tags) and to show the strength between nodes. This will result in accurate clustering of tags. In addition, this approach can manage a large part of tags in this way. To obtain semantic relevance, the property of co-occurrence tags is used. However, unlike the previous methods, it takes into account the fact that users' interests change over time.

As a result, when using the co-occurrence tags, it added the time of tag assignment parameter. If the two co-occurrence tags are close to each other in the parameter, they will have a higher score and therefore, the power of semantic correlation will be higher. Jaccard similarity is used to find similarity of co-occurred tags. The formula is defined as follow:

$$sim_{Jac}(t_i, t_j) = \frac{|R(t_i) \cap R(t_j)|}{|R(t_i) \cup R(t_j)|} \quad (1)$$

where $R(t_i)$ stands for the set of resources tagged by the $t_i$ tag. When two tags co-occur, first their semantic similarity is calculated with the Jaccard similarity formula. Then the lexical similarity with Levenshtein distance is calculated and called $sim_{Lev}$. For calculating lexical relevance and morphological tags, we used high-threshold $sim_{Lev}$ criterion. This can resolve minor morphological changes as well as misspellings (these are two common problems with social tagging systems). Moreover, for tags that do not have semantic relevance but they have a strong lexical similarity, this lexical similarity is considered as the weights of the edges. The formula for $sim_{Lev}$ is defined as follow:

$$sim_{Lev} = \frac{1 - Levenshtein\ distance(t_i, t_j)}{maxLength(t_i, t_j)} \quad (2)$$

After obtaining both similarities, each lexical or semantic similarity that is larger, selected as the similarity between two tags. If two tags do not co-occur and the lexical similarity is greater than a threshold value of α, then the value is selected as a similarity between two tags. Since the users' interests change over time, we considered another similarity based on the time of tag assignments for co-occurred tags. This similarity is shown with $sim_{time}(t_i,t_j)$. Suppose that Timestamp($t_j$, $r_k$) shows the last time the tag $t_i$ is assigned to the resource $r_k$. The set of the common resources for two co-occurred tags $t_i$ and $t_j$, whose assignment is too close, is shown by nco($t_i,t_j$). The formulas of $nco(t_i, t_j)$ and $sim_{time}(t_i,t_j)$ can be defined as follows:

$$\mathrm{nco}(t_i, t_j) = \{r_k | r_k \in R(t_i) \land r_k \in R(t_j) \land |\mathrm{Timestamp}(t_i, r_k) - \mathrm{Timestamp}(t_j, r_k)| \leq \tau\} \quad (3)$$

$$sim_{time}(t_i, t_j) = \frac{|\#nco(t_i,t_j)|}{|R(t_i) \cap R(t_j)|} \quad (4)$$

Therefore, $sim_{time}$ is considered when two tags are co-occurred. Finally, a graph of the tags is created and the weight between two desired nodes calculated by Eq. 5.

$$w((t_i, t_j) = \begin{cases} \lambda \times sim_{jac}(t_i, t_j) + (1-\lambda)\ sim_{time} & if\ t_i\ co-occurred\ with\ t_j\ and\ sim_{jac} > sim_{lev} \\ \lambda \times sim_{lev}(t_i, t_j) + (1-\lambda)\ sim_{time} & if\ t_i\ co-occurred\ with\ t_j\ and\ sim_{jac} < sim_{lev} \\ \lambda \times sim_{lev} & if\ t_i\ not\ cooccurred\ with\ t_j\ and\ sim_{lev} > threshold \\ 0 & otherwise \end{cases} \quad (5)$$

The pseudocode for generating the graph of tags is shown in Algorithm 1.

Algorithm 1: Generating the Graph of tags

G= GenGraph(T)
Input: T: The set of tags of dataset
Output: G= (V, E), V are nodes and E are weighted edges in G
V={}

```
E={}
for each tᵢ∈ T do
    for each tⱼ∈ T do
        Calculate w(tᵢ,tⱼ) according to Eq. 5
        Add tᵢ , tⱼ to V if not exist.
        Add an edge between tᵢ and tⱼ with weight w(tᵢ,tⱼ ) to E
    end for
end for
```

After generating a graph of tags, their communities are determined. In the following, the community detection algorithms and the reasons for using them in the proposed method will be explained.

### 3.2. Community detection

The scope of social networks is known as a significant evolution in the last decade, and community detection has emerged to analyze many fields as well as the individual's interactions within social environments [41]. In this work, it has been decided to use this method to analyze tags. Therefore, the second stage of the first phase of the proposed approach is to detect tag communities. The best way to analyze the network of tags and to cluster them, is to use community detection methods. The purpose of detecting communities is to extract groups whose internal communications of their communities are stronger and more powerful than external communications. In fact, with this method, the existing divisions in a network can be identified and separated to get a better view of the structure of a network for its analysis. Various methods have been proposed for community detection. Here, community means a group of network nodes of tags that are tightly connected. The strength of the joints is obtained through their degree of similarity. In other words, the strength of the power connections shows the semantic and lexical similarity. In fact, the nodes belonging to the same community are similar and related to the same interest. The better the identified communities, the more accurate results are obtained in the recommendation section of the research system. Here, the criterion for distinguishing a good community is modularity, which is widely used in community detection methods. Modularity is defined based on [42] in Eq. 6.

$$Q = \frac{1}{2m} \sum_{ij} \left( w(t_i, t_j) - \frac{K_{t_i} K_{t_j}}{2m} \right) \delta\left(Cm_{t_i}, Cm_{t_j}\right) \tag{6}$$

$$K_{t_i} = \sum_k w(t_i, t_k) \tag{7}$$

$$m = \frac{1}{2} \sum_{ij} w(t_i, t_j) \tag{8}$$

Where w(tᵢ,tⱼ) is the weight of two nodes tᵢ and tⱼ. In the following $K_{t_i}$ is the degree of node tᵢ defined in Eq. 7. Also, $\delta\left(Cm_{t_i}, Cm_{t_j}\right)$ has a value of 1 if both nodes tᵢ and tⱼ belong to the same community; otherwise, its value is zero. In Eq. 6, m is the total weights of all edges in this graph, defined in Eq.8. In this research, the Louvain method has been used to identify the tags community. This method is non-overlapping. After this step, the next part, which is the presentation of resource recommendation stage, will be explained.

### 3.3. Resources recommendation stage

In this Section, after identifying the communities, the recommendation steps are explained. First C is defined as a set of communities which is detected by the community detection algorithm. Each community is a set of tags. These are defined in Eq.9 and Eq. 10.

$$C = \{c_i | c_i \text{ is a community}\} \tag{9}$$

$$c_i = \{t_k | t_k \in T, \theta(t_k) = i\} \tag{10}$$

where θ stands for a function determining the community number of tags. In the following, for each resource a probability value is calculated for each community that indicates the probability of that resource's membership in the desired community of tags which can be defined by Eq. 11.

$$Pr(r_i, c_j) = \frac{N(r_i, c_j)}{\sum_{c_j \in C} N(r_i, c_j)} \tag{11}$$

$$N(r_i, c_j) = \sum_{t_k \in c_j} N(t_k, r_i, c_j) \tag{12}$$

In Eq. 12, $N(t_k, r_i, c_j)$ is the number of tags are used in the community $c_j$ to be tagged the resource $r_i$, where it is possible to determine which communities are related to the resource. The higher the probability, the more relevant the resource is to that community. In other words, more tags from a community are used to tag the resource. In fact, by examining a user's resources, it is possible to determine the user's interests in various communities.

In this research, it was determined experimentally that the overlap of resource communities is high. The creation of resource communities through tag communities causes this high percentage. Due to the reduced accuracy of the recommendations. Therefore, at this stage, the resource communities are refined. According to Eq. 11, the probability value a resource to a desired community is obtained, which we can consider threshold value. Therefore, resources that are less than threshold dependent are excluded from that community. In this way, the resulting communities will be more reasonable.

After determining the resources' membership for different communities, in the next step, by Eq. 13, which is the Ellenberg similarity criterion, the degree of similarity between two resources $r_i$ and $r_j$ is obtained.

$$Sim_e(r_i, r_j) = \frac{m/2}{m/2 + b + c} \tag{13}$$

where m is the sum of the probabilities of membership the two resources $r_i$ and $r_j$ in the common communities. b indicates the probability of membership the resource $r_i$ and c represents the probability of membership the resource $r_j$ in different communities. By calculating the similarity between resources of the target user and resources of the specified communities of the target user, a list of recommended candidate resources can be obtained. More formally, let $R_{TU}$ and $R_{TC}$ be the set of the target user resources and resources of the target user communities. This list has two problems. The first problem is that there are too many resources in this list that have the same amount of similarity to the resources of the target user, and it is difficult to choose the exact resources recommended and close to the user's interest. The second problem is that there are many unrelated resources to a reasonable degree of similarity in this list, if numerical similarity is sufficient, the expected result will not be obtained. To solve this problem, the other similarity has been used between two resources by Eq.14.

$$Sim_u(r_i, r_j) = |(U(r_i) \cap U(r_j))| / \max\{|U(r_i)|, |U(r_j)|\} \tag{14}$$

$U(r_i)$ stands for the set of users that annotated $r_i$ with tags. Then by Eq. 15, resources with the most similarity to the resources of the target user are calculated.

$$\text{Msr}(r_i) = \underset{r_i}{argmax}\{ \text{Sim}_u(r_i, r_j) + \text{Sim}_e(r_i, r_j) \} \quad (15)$$

where $r_i$ is the resource of the target user ($r_i \in R_{TU}$) and $r_j$ is the resource of the target community ($r_j \in R_{TC}$). Finally, a list of recommended resources is obtained by Eq. 16.

$$\text{Recommend list} = \{\text{Msr}(r_i) \mid r_i \in R_{TU}\} \quad (16)$$

The pseudocode for generating the recommended list is shown in Algorithm 2.

---

**Algorithm 2: generating the recommended list**

---

R_List = **Generate_recommended_list (n_com, $R_{TU}$, $R_{TC}$, C)**
**Input:** n-com = number of communities, $R_{TU}$: set of the target user resources,
$R_{TC}$: set of resources of the target communities, C={$c_i$|$c_i$ is community of the target user}
**Output:** R_List: list of recommended resources for the target user.
L= {}
R_List= {}
for each $r_i \in R_{TC}$
  for each $c_j \in C$
    calculate $Pr(r_i, c_j)$ according to Eq.11 and Eq. 12
    if $(Pr(r_i, c_j) \neq 0)$ then
      A(i,j)= $Pr(r_i, c_j)$
  end for
end for
for each $r_i \in R_{TU}$
  for each $r_j \in R_{TC}$
    m=0; b=0; c=0;
    for each k $\in$ n_com
      if (A(i, k) $\neq$ 0 And A(j, k) $\neq$ 0) then m=m+ $A(i,k) + A(j,k)$
      else if (A(i, k) $\neq$ 0) then b=b+ $A(i,k)$
      else c=c+ $A(j,k)$
    $\text{Sim}_e(r_i, r_j) = \frac{m/2}{m/2+b+c}$
    calculate $\text{Sim}_u(r_i, r_j)$ according to Eq.14
    append(L, $r_i$, $r_i$, $\text{Sim}_u(r_i, r_j)$+ $\text{Sim}_e(r_i, r_j)$)
  end for
end for
  calculate Msr($r_i$) from L according to Eq. 15
  append(R_List , Msr($r_i$))
return (R_List)

---

## 4. Performance evaluation of proposed solution

### 4.1. Experimental dataset

One of the main parts of each recommendation system is the collection of information. If it were done in a regular and accurate manner, the analysis of data will be accomplished with great speed and accuracy [43]. In the proposed method, among the valid datasets that have been published, two datasets were used.

1) **Del.icio.us**: the highly used Hetrec2011-Del.icio.us-2k dataset by Zuo rt al. [44]and Xu et al. [45] in the experiments, which includes 53,388 tags, 1867 users, and 69,226 sources, which are gathered from Del.icio.us.com and released in [46]. In this dataset, users not only can save and organize their favorite pages (URLs) but also tag and share them as they wish. Users are connected in a social network created from Del.icio.us interactions, and each user has its own tags, bookmarks and tag assignments.
2) **Last.Fm:** this is an artist recommendation dataset and gathered from music system Last.Fm[2], which users are able to tag artists. Therefore, each user to artists has a list of tag assignments [47]. This dataset includes 11946 tags, 1892 users and 17632 artists.

In the beginning of using these datasets, we first removed the noisy and meaningless tags. Since there are some special characters and numbers in the dataset, these tags are noisy and meaningless. In the beginning of using the dataset, these tags have been removed and the data set has been cleaned. Python scripts have been used to perform the cleaning dataset.

Unlike most previous methods, the tags are used with any number of repetitions in this work. Therefore, the proposed solution is responsive to the cold start problem. Then, the test and trained dataset are specified as 20% and 80% of the total data. Recommendations are generated based on the known information in the training set, and then the test set is used to evaluate the performance of recommendation algorithms

### 4.2. Experimental parameters and baseline methods

In the first step of the proposed method, which is generating a graph tags, a graph of all user tags was created. In creating this graph, Jaccard similarity and Levenshtein distance was used to determine the edge weight between two tags for co-occurring tags. In order to consider the lexical connection between them, Levenshtein distance was used and the greatest similarity was selected. If two tags co-occur, also the time of tag assignments are considered. For tags that do not co-occur, the lexical similarity ($sim_{Lev}$) to α threshold is supposed. Here, in various experiments, we experimentally considered the lexical similarity threshold, α equal to 0.7 for co-occur tags; otherwise, its value is 0.8. If the lexical similarity were greater than this threshold, it calculated as 50% and applied as weight.

To show the efficiency of the proposed method, this method is compared with the following models:

1) CCS[3] method: The Cosine similarity method is based on clustering. Hierarchical clustering [48] was used to model users and resources as a vector of cluster-based attributes, and content-based filtering is based on cosine similarity of recommendations. The proposed method is better than this method for several reasons. First, use the tags graph and create this graph in a powerful way. Secondly, the use of robust graph analysis, which is a method of community detection. The results of these two methods show the superiority of this research method.

2) ACF[4] method: uses the CF method based on automatic encoder. An automated encoder is usually used to obtain summary introductions from user profiles based in which CF recommendations are used. Experiments on CF method with different number of hidden layers demonstrate that deeper architectures can work better if the depth of the neural network is set appropriately [45].

3) CCF[5] method or CF based on clustering: It is similar to CCS method but here the user-based CF method is used for recommendations [49].

4) PMF[6] method: This technique, which is based on filtering user collaboration, uses a user ranking matrix. This model, based on the assumption that users who have rated similar sets of items are likely to have similar preferences [50]. The method was chosen to demonstrate the superiority of using another dimension of information, namely tags.

---

[2] 2http://www.last.fm.com
[3] Clustering-based Cosine similarity
[4] Autoencoder-based collaborative filtering
[5] Clustering-based collaborative filtering
[6] Probabilistic Matrix Factorization

These two criteria have been significantly improved in the proposed algorithm, according to the known algorithms. The results of comparing the presented method with the proposed and known methods are shown in Table 1.

5) KGAT: this is state-of-the-art knowledge-based model, performs knowledge-aware attentive graph convolution in KG for high-order modeling of relation [51].

### 4.3. Evaluation metrics

The criteria most often used to evaluate recommender systems are Precession (P), Recall (R) and all of which used to evaluate the quality performance of recommendation systems. In fact, the criterion of Precession(P) determines what percentage of the set of recommenders is presented by a method is correct. This criterion measures the correctness and accuracy of the proposals recommended, as a result, the larger the criterion, the less errors in the method being measured. The next criterion, which is Recall(R), refers to what percentage of the offers are really users' interest. According to [26], P and R, are defined in Eq. 17 and Eq. 18. Since users usually review the highest recommended items, we cut these criteria to a specific rank k. That is, just considering k the number of results at the top of the recommendation list, the precession in k with P@k and recall in k by R@k.

$$P = \frac{|rr \cap tr|}{|rr|} \quad (17)$$

$$R = \frac{|rr \cap tr|}{|tr|} \quad (18)$$

The experiments, the mean values of P@k and R@k were used to evaluate the performance of the system recommended by users. Where rr is the list of recommended resources and tr is the list of resources being tested. The higher value of these two criteria in different methods indicates their better quality. All experiments are implemented an Intel(R) Core i7 computer with 2.67 GHz CPU and 16.00 GB of RAM.

### 4.4. Experimental results and analysis

Two research questions have been raised in this section, and the experiments were designed to address these questions.

RQ1: How effective is the lexical similarity in the proposed method? To answer this RQ1, the proposed method has been compared with lexical similarity which is shown as LEXSEM_CDR. The method without this similarity is demonstrated by SEM_CDR. We examined them with using two metrics, Recall@k and Precision@k with four k values, 5, 10, 15 and 20. Table 1. represents the experimental results. These results show that lexical similarity enhances system performance by using semantic similarity. This improvement in output is especially evident when there are spelling mistakes in the tags. In these experiments, we experimentally considered the threshold value to be 0.7 for co-occur tags; otherwise, its value is 0.8. Because in assigning tags to resources, spelling mistakes are obvious, instead of spending time to clean them, using lexical similarity seemed very useful.

Table 1: Recommendation Performance on Del.icio.us dataset for two Proposal Models (in %)

| Models | P@5 | P@10 | P@15 | P@20 | R@5 | R@10 | R@15 | R@20 |
|---|---|---|---|---|---|---|---|---|
| LEXSEM_CDR | 24.08 | 22.80 | 21.70 | 20.73 | 8.11 | 14.60 | 18.84 | 25.54 |
| SEM_CDR | 24.02 | 22.67 | 21.52 | 20.07 | 8.06 | 14.03 | 18.44 | 25.32 |

Table 2: Recommendation Performance on Last.Fm dataset for two Proposal Models (in %)

| Models | P@5 | P@10 | P@15 | P@20 | R@5 | R@10 | R@15 | R@20 |
|---|---|---|---|---|---|---|---|---|

| LEXSEM_CDR | 17.09 | 16.60 | 15.40 | 12.80 | 7.45 | 8.89 | 10.89 | 12.66 |
| SEM_CDR | 16.95 | 16.59 | 15.40 | 12.80 | 7.32 | 8.89 | 10.86 | 12.66 |

Tables 1, 2 show the results that "Del.icio.us" dataset performs well compared to "Last.Fm" due to the diversity in tags by applying two similarities.

In the following, the second question will be raised.

RQ2: How effective is the time of tag assignment in the proposed methods? Due to the changes in users' interests over time, we have considered the time difference of the last assignment of two co-occurred tags to generate a graph. Tables 3,4 present the results of the proposed method (CDR_TIME) and its comparison with LEXSEM_CDR. The results show that using the time of tag assignment approved accuracy of the recommender system.

Table 3 : Recommendation Performance on Del.icio.us dataset for two Proposal Models (in %)

| Models | P@5 | P@10 | P@15 | P@20 | R@5 | R@10 | R@15 | R@20 |
|---|---|---|---|---|---|---|---|---|
| LEXSEM_CDR | 24.08 | 22.80 | 21.70 | 20.73 | 8.11 | 14.60 | 18.84 | 25.54 |
| CDR_TIME | 24.23 | 22.98 | 21.87 | 20.89 | 8.43 | 14.84 | 19.21 | 25.96 |

Table 4: Recommendation Performance on Last.Fm dataset for two Proposal Models (in %)

| Models | P@5 | P@10 | P@15 | P@20 | R@5 | R@10 | R@15 | R@20 |
|---|---|---|---|---|---|---|---|---|
| LEXSEM_CDR | 17.09 | 16.60 | 15.40 | 12.80 | 7.45 | 8.89 | 10.89 | 12.66 |
| CDR_TIME | 18.18 | 17.26 | 17.96 | 15.98 | 7.98 | 13.001 | 13.21 | 14.18 |

In "Last.Fm" dataset, users' tastes vary over time, so by applying the time parameter, compared to dataset " Del.icio.us" , the results change noticeably.

The results in tables 5,6 show that the proposed method has significantly improved the two criteria of precision and recall. In addition to the improved results, the great advantage of the proposed method is that it uses only training data and does not use any external knowledge base or resource contents. The proposed method is agile and simple. As it is explained in the previous sections, tags do not have a specific format and users choose them without restrictions and this was the main reason for not using an external knowledge.

On the other hand, the reason why KGAT works more accurately with the increase in the number of recommendations is that, the latent relations are better extracted in this method. At the same time, the simplicity of the proposed method is considered an advantage over the KGAT method.

Table 5: Comparisons of Performance on Del.icio.us dataset for different models (in %)

| Models | p@5 | p@15 | p@30 | p@50 | R@5 | R@15 | R@30 | R@50 |
|---|---|---|---|---|---|---|---|---|
| CCF | 0.913 | 0.757 | 0.597 | 0.454 | 0.439 | 1.051 | 1.499 | 1.803 |
| ACF | 1.120 | 0.909 | 0.736 | 0.595 | 0.590 | 1.209 | 1.917 | 2.364 |
| CCS | 2.397 | 1.903 | 1.564 | 1.273 | 0.938 | 2.271 | 3.774 | 4.774 |
| PMF | 9.157 | 7.467 | 6.784 | 6.306 | 1.302 | 2.851 | 4.988 | 7.587 |
| KGAT | 24.03 | 21.18 | **19.80** | **18.02** | 8.13 | 19.18 | **27.98** | **37.86** |
| **CDR(proposed)** | **24.23** | **21.87** | 19.75 | 17.80 | **8.43** | **19.21** | 27.92 | 37.45 |

Table 6: Comparisons of Performance on Last,Fm dataset for different models (in %)

| Models | p@5 | p@15 | p@30 | p@50 | R@5 | R@15 | R@30 | R@50 |
|---|---|---|---|---|---|---|---|---|
| CCF | 1.004 | 0.987 | 0.765 | 0.668 | 0.439 | 1.051 | 1.499 | 1.803 |
| ACF | 1.820 | 1.004 | 0.845 | 0.702 | 0.590 | 1.209 | 1.917 | 2.364 |
| KGAT | 18.08 | **17.98** | **16.95** | **15.13** | 7.54 | **16.25** | **24.13** | **33.96** |
| **CDR(proposed)** | **18.18** | 17.96 | 14.65 | 12.65 | **7.98** | 13.21 | 14.95 | 19.75 |

### 4.6. Threats to validity

In this section, a brief list of internal and external threats related to the validity of research findings after various reviews is provided. The internal threats to the validity of the findings this research: The accuracy of the proposed method depends on quality of input dataset. In this work, after removing the noisy and meaningless data, all the data are divided into training and test data and the proposed method has been applied to it. By examining the results obtained on two datasets and comparing them with the state-of-the-art methods, it was concluded that this method has worked well. The values of the parameters used in the calculation of the similarity of the tags have been tried to select the best value by performing various tests, and the results of some of them have been randomly selected and checked manually. This review has confirmed the correctness of choosing the appropriate values.

The external threats to the validity of the findings this research: The ability to generalize the algorithm is one of the external threats. Therefore, the proposed method has been tested on two datasets. The results in both datasets confirm the superiority of this method. As a result, this method can be used in similar datasets without problems. The algorithm's performance may depend on the quality and size of the dataset used to train and test the algorithm. If the dataset is biased or not representative, the resulting may not accurately reflect the relationships between tags, users, and resources. And this will affect the accuracy of the recommendation system. As usual, here the data is selected as 80% and 20% for training and testing, respectively. Tag usage behavior may change over time or in response to different situations, which may affect the accuracy of the resulting. To deal with this threat in the proposed method, we used the time parameter to consider the user's behavioral changes during different times. The datasets used are static and their dynamicity needs further investigation, which was not specifically investigated in this research.

### 5. Conclusion and future works

In the method, the problems in tagged recommender systems and concluded that the performance of these systems is affected by semantic and lexical ambiguities. Various solutions have been proposed in this field, most of which suggested the use of external knowledge and thesaurus. Because the use of tags in some cases does not follow any specific rules, these solutions were not suitable especially in datasets such as Del.icio.us, where most tags are not covered by thesaurus. Therefore, by using co-occurring tags, the time of tag assignment and statistical and mathematical methods, we identified the semantic and lexical similarity of more accurate communications. The proposed method reached a suitable modeling of users' interest from the community detection method. Based on this accurate modeling it achieved better results in providing recommendations. The results of experimental examinations also confirmed this. The performance of the proposed method has been done using two criteria of precision and recall based on evaluations with "Del.icio.us" and "Last.Fm" dataset. The evaluation results show that the precision and recall of the proposed method have significantly improved, compared to the other methods. According to the experimental results, the criteria of recall and precision have been improved, on average by 5% and 7% respectively. In later studies, we plan to use more advanced community detection methods to cluster tags and get more accurate results and also will provide a plan to eliminate the semantic ambiguity of the tags.

**REFERENCES**

1. S. Zhang, L. Yao, A. Sun, and Y. Tay, "Deep learning based recommender system: a survey and new perspectives," ACM Computing Surveys, vol. 52, no. 1, pp. 1–38, 2019.
2. J. Bobadilla, F. Ortega, A. Hernando, and A. Gutiérrez, "Recommender systems survey," Knowledge-Based Systems, vol. 46, pp. 109–132, 2013.
3. N. Nikzad-Khasmakhi, M. A. Balafar, M. R. Feizi-Derakhshi, and C. Motamed, "BERTERS: multimodal representation learning for expert recommendation system with transformers and graph embeddings," Chaos, Solitons & Fractals, vol. 151, p. 111260, 2021.
4. P. Zhang, D. Wang, and J. Xiao, "Improving the recommender algorithms with the detected communities in bipartite networks," Physica A: Statistical Mechanics and its Applications, vol. 471, pp. 147–153, 2017.
5. Y. Shi, M. Larson, and A. Hanjalic, "Collaborative filtering beyond the user-item matrix: a survey of the state of the art and future challenges," ACM Computing Surveys, vol. 47, no. 1, pp. 1–45, 2014.
6. A. Mathes, "Folksonomies–cooperative classification and communication through shared metadata," LIS590CMC (Doctoral Semin. Grad. Sch. Libr. Inf. Sci. Univ. Illinois Urbana-Champaign, Accessed: Dec. 26, 2020.
7. A. S. Ghabayen and S. A. M. Noah, "Using tags for measuring the semantic similarity of users to enhance collaborative filtering recommender systems," International Journal on Advanced Science, Engineering and Information Technology, vol. 7, no. 6, pp. 2063–2070, 2017.
8. Q. Qi, Z. Chen, J. Liu, C. Hui, and Q. Wu, "Using inferred tag ratings to improve user-based collaborative filtering," in Proceedings of the 27th Annual ACM Symposium on Applied Computing, pp. 2008–2013, Association for Computing Machinery, March, 2012.
9. H. Liang, Y. Xu, Y. Li, and R. Nayak, "Tag based collaborative filtering for recommender systems," in Rough Sets and Knowledge Technology, P. Wen, Y. Li, L. Polkowski, Y. Yao, S. Tsumoto, and G. Wang, Eds., pp. 666–673, Springer, Berlin, Heidelberg, 2009.
10. H. Wu, Y. Pei, B. Li, Z. Kang, X. Liu, and H. Li, "Item recommendation in collaborative tagging systems via heuristic data fusion," Knowledge-Based Systems, vol. 75, pp. 124–140, 2015.
11. S. Goel and R. Kumar, "Folksonomy-based user profile enrichment using clustering and community recommended tags in multiple levels," Neurocomputing, vol. 315, pp. 425–438, 2018.
12. V. Zanardi and L. Capra, "A scalable tag-based recommender system for new users of the social web," in Database and Expert Systems Applications, A. Hameurlain, S. W. Liddle, K. D. Schewe, and X. Zhou, Eds., pp. 542–557, Springer, Berlin Heidelberg, 2011.
13. G. Solskinnsbakk and J. A. Gulla, "Mining tag similarity in folksonomies," in Proceedings of the 3rd International Workshop on Search and Mining User-generated Contents, pp. 53–60, Association for Computing Machinery, October, 2011.
14. E. Djuana Tjhwa, Y. Xu, and Y. Li, "Constructing tag ontology from folksonomy based on WordNet," in Proceedings of the IADIS International Conference on Internet Technologies and Society 2011, pp. 85–93, IADIS Press, 2011.

15. A. García-Silva, L. J. García-Castro, A. García, and O. Corcho, "Social tags and linked data for ontology development: a case study in the financial domain," in Proceedings of the 4th International Conference on Web Intelligence, Mining and Semantics (WIMS14), pp. 1–10, Association for Computing Machinery, June, 2014.
16. J. Chen, S. Feng, and J. Liu, "Topic sense induction from social tags based on non-negative matrix factorization," Information Sciences, vol. 280, pp. 16–25, 2014.
17. K. H. L. Tso-Sutter, L. B. Marinho, and L. Schmidt-Thieme, "Tag-aware recommender systems by fusion of collaborative filtering algorithms," in Proceedings of the 2008 ACM symposium on Applied computing, pp. 1995–1999, Association for Computing Machinery, March, 2008.
18. S. Niwa, T. Doi, and T. S. Honiden, "Web page recommender system based on folksonomy mining for ITNG' 06 submissions," in Third International Conference on Information Technology: New Generations (ITNG'06), pp. 388–393, IEEE, April, 2006.
19. S. Sen, J. Vig, and J. Riedl, "Tagommenders: connecting users to items through tags," in Proceedings of the 18th International Conference on World Wide Web, pp. 671–680, Association for Computing Machinery, 2009, April.
20. X. Zheng, M. Wang, C. Chen, Y. Wang, and Z. Cheng, "EXPLORE: EXPLainable item-tag CO-REcommendation," Information Sciences, vol. 474, pp. 170–186, 2019.
21. G. W. Furnas, C. Fake, L. von Ahn, J. Schachter, S. Golder, K. Fox, and M. Naaman, "Why do tagging systems work?" in CHI'06 Extended Abstracts on Human Factors in Computing Systems, pp. 36–39, Association for Computing Machinery, April, 2006.
22. P. Lamere, "Social tagging and music information retrieval," Journal of New Music Research, vol. 37, no. 2, pp. 101–114, 2008.
23. C. Biancalana and A. Micarelli, "Social tagging in query expansion: a new way for personalized web search," in 2009 International Conference on Computational Science and Engineering, pp. 1060–1065, IEEE, Vancouver, BC, Canada, August, 2009.
24. S. A. Golder and B. A. Huberman, "Usage patterns of collaborative tagging systems," Journal of Information Science, vol. 32, no. 2, pp. 198–208, 2006.
25. Z. Xu, Y. Fu, J. Mao, and D. Su, "Towards the semantic web: collaborative tag suggestions," in Collaborative web tagging workshop at WWW2006, Edinburgh, Scotland, May, 2006.
26. J. Zhang, Q. Peng, S. Sun, and C. Liu, "Collaborative filtering recommendation algorithm based on user preference derived from item domain features," Physica A: Statistical Mechanics and its Applications, vol. 396, pp. 66–76, 2014.
27. Y. Wu, Y. Yao, F. Xu, H. Tong, and J. Lu, "Tag2word: Using tags to generate words for content based tag recommendation," in Proceedings of the 25th ACM International on Conference on Information and Knowledge Management, pp. 2287–2292, Association for Computing Machinery, October, 2016.
28. M. de Gemmis, P. Lops, G. Semeraro, and P. Basile, "Integrating tags in a semantic content-based recommender," in Proceedings of the 2008 ACM Conference on Recommender Systems, pp. 163–170, Association for Computing Machinery, October, 2008.
29. C. Wartena, R. Brussee, and M. Wibbels, "Using tag co-occurrence for recommendation," in 2009 Ninth International Conference on Intelligent Systems Design and Applications, pp. 273–278, IEEE, Pisa, Italy, November, 2009.
30. D. Kowald, S. Kopeinik, P. Seitlinger, T. Ley, D. Albert, and C. Trattner, "Refining frequency-based tag reuse predictions by means of time and semantic context," in Mining, Modeling, and Recommending 'Things' in Social Media, M. Atzmueller, A. Chin, C. Scholz, and C. Trattner, Eds., pp. 55–74, Springer, Cham, 2013.
31. D. I. Ignatov, S. I. Nikolenko, T. Abaev, and J. Poelmans, "Online recommender system for radio station hosting based on information fusion and adaptive tag-aware profiling," Expert Systems with Applications, vol. 55, pp. 546–558, 2016.

32. A. Vall, "Listener-inspired automated music playlist generation," in Proceedings of the 9th ACM Conference on Recommender Systems, pp. 387–390, Association for Computing Machinery, September, 2015.
33. E. Zheng, G. Y. Kondo, S. Zilora, and Q. Yu, "Tag-aware dynamic music recommendation," Expert Systems with Applications, vol. 106, pp. 244–251, 2018.
34. H. Xie, X. Li, T. Wang, R. Y. K. Lau, T.-L. Wong, L. Chen, F. L. Wang, and Q. Li, "Incorporating sentiment into tag-based user profiles and resource profiles for personalized search in folksonomy," Information Processing & Management, vol. 52, no. 1, pp. 61–72, 2016.
35. Fernández-Tobías and I. Cantador, "Exploiting social tags in matrix factorization models for cross-domain collaborative filtering," in CBRecSys@ RecSys, pp. 34–41, October, 2014.
36. H. Liu, "Resource recommendation via user tagging behavior analysis," Cluster Computing, vol. 22, no. S1, pp. 885–894, 2019.
37. F. Jabeen, S. Khusro, A. Majid, and A. Rauf, "Semantics discovery in social tagging systems: a review," Multimedia Tools and Applications, vol. 75, pp. 573–605, 2016.
38. J. Beel, B. Gipp, S. Langer, and C. Breitinger, "Research-paper recommender systems: a literature survey," International Journal on Digital Libraries, vol. 17, pp. 305–338, 2016.
39. R. Jäschke, A. Hotho, C. Schmitz, B. Ganter, and G. Stumme, "Discovering shared conceptualizations in folksonomies," Journal of Web Semantics, vol. 6, no. 1, pp. 38–53, 2008.
40. M. Ahmed, M. D. Ansari, N. Singh, V. K. Gunjan, B. V. Santhosh Krishna, and M. Khan, "Rating-based recommender system based on textual reviews using iot smart devices," Mobile Information Systems, vol. 2022, Article ID 2854741, 18 pages, 2022.
41. S. Souabi, A. Retbi, M. K. Idrissi, and S. Bennani, "Toward a recommendation-oriented approach based on community detection within social learning network," in Advanced Intelligent Systems for Sustainable Development (AI2SD'2019), M. Ezziyyani, Ed., pp. 217–229, Springer, Cham, 2020.
42. M. E. J. Newman and M. Girvan, "Finding and evaluating community structure in networks," Physical Review E, vol. 69, no. 2, Article ID 026113, 2004.
43. N. Nikzad-Khasmakhi, M. A. Balafar, and M. Reza Feizi-Derakhshi, "The state-of-the-art in expert recommendation systems," Engineering Applications of Artificial Intelligence, vol. 82, pp. 126–147, 2019.
44. Y. Zuo, J. Zeng, M. Gong, and L. Jiao, "Tag-aware recommender systems based on deep neural networks," Neurocomputing, vol. 204, pp. 51–60, 2016.
45. Z. Xu, C. Chen, T. Lukasiewicz, Y. Miao, and X. Meng, "Tag-aware personalized recommendation using a deep-semantic similarity model with negative sampling," in Proceedings of the 25th ACM International on Conference on Information and Knowledge Management, pp. 1921–1924, Association for Computing Machinery, October, 2016.
46. I. Cantador, P. Brusilovsky, and T. Kuflik, "Second workshop on information heterogeneity and fusion in recommender systems (HetRec2011)," in Proceedings of the Fifth ACM Conference on Recommender Systems, pp. 387-388, Association for computing machinery, October, 2011.
47. X. Wang, X. He, Y. Cao, M. Liu, and T.-S. Chua, "KGAT: Knowledge graph attention network for recommendation," in Proceedings of the 25th ACM SIGKDD International Conference on Knowledge Discovery & Data Mining, pp. 950–958, Association for computing machinery, July, 2019.
48. A. Shepitsen, J. Gemmell, B. Mobasher, and R. Burke, "Personalized recommendation in social tagging systems using hierarchical clustering," in Proceedings of the 2008 ACM Conference on Recommender Systems, pp. 259–266, Association for computing machinery, October, 2008.
49. Z. Xu, D. Yuan, T. Lukasiewicz, C. Chen, Y. Miao, and G. Xu, "Hybrid deep-semantic matrix factorization for tag-aware personalized recommendation," in ICASSP 2020-2020 IEEE International Conference on Acoustics, Speech and Signal Processing (ICASSP), pp. 3442–3446, IEEE, Barcelona, Spain, May, 2020.

50. R. Salakhutdinov and A. Mnih, "Probabilistic matrix factorization advances in neural information processing systems 21 (NIPS 21)," Vancouver, Canada, 2008.
51. X. Wang, X. He, Y. Cao, M. Liu, and T.-S. Chua, "KGAT: Knowledge graph attention network for recommendation," in Proceedings of the 25th ACM SIGKDD International Conference on Knowledge Discovery & Data Mining, pp. 950–958, Association for computing machinery, July, 2019.